\documentclass[12pt,thmsa]{article}
\usepackage{sw20lart}

\input{tcilatex}

\begin{document}

\author{Fayyazuddin \\
%EndAName
National\ Centre for Physics and Department of Physics\\
Quaid-i-Azam University\\
Islamabad, Pakistan.}
\title{$SU(3)$ and $CP$ violating weak and strong final state phases for $%
B_{d}^{0}$ and $B_{s}^{0}$ decays}
\maketitle

\begin{abstract}
Using rotation in $SU(3)$ space, a set of relations between various decay
modes of $B_{d}$ and $B_{s}$ are derived. The decays $\bar{B}%
_{d}^{0}\rightarrow K^{*-}\pi ^{+}(\rho ^{+}K^{-}),$ $\bar{B}%
_{s}^{0}\rightarrow K^{*-}K^{+}$ are expressed in terms of decay parameters
of $\bar{B}_{d}^{0}\rightarrow \rho ^{-}\pi ^{+}(\rho ^{+}\pi ^{-})$. In
particular the parameters $r_{-+}(r_{+-})$ of $B_{d}\rightarrow \rho \pi $
decays are obtained in terms of experimentally known decay rates $%
R_{-+}(R_{+-})=\frac{1}{2}\left( \Gamma _{\rho ^{+}\pi ^{-}(\rho ^{-}\pi
^{+})}+\bar{\Gamma}_{\rho ^{-}\pi ^{+}(\rho ^{+}\pi ^{-})}\right) ,$ $%
R_{-+}^{\prime }(R_{+-}^{\prime })=\frac{1}{2}\left( \Gamma _{K^{*+}\pi
^{-}(\rho ^{-}K^{+})}+\bar{\Gamma}_{K^{*-}\pi ^{+}(\rho ^{+}K^{-})}\right) ,$
known parameters $\bar{\lambda},$ $f_{K^{*}}/f_{\rho }(f_{K}/f_{\pi })$ and
two parameters $B_{-+}=\frac{R_{-+}}{\left| T^{-+}\right| ^{2}},B_{+-}=\frac{%
R_{+-}}{\left| T^{+-}\right| ^{2}}$ which are determined by using
factorization for tree amplitudes $T^{-+}$ and $T^{+-}$. We find $%
r_{-+}=0.21\pm 0.04,r_{+-}=0.25\pm 0.06.$ With these values the following
bounds on $\left( z\equiv \cos \gamma \cos \delta ,\text{ }x=\sin \gamma
\sin \delta \right) $ are derived: 
\[
-0.34(-0.33)\leq z_{-+}(z_{+-})\leq 0.28(0.27) 
\]

and 
\[
0.16(0.43)\leq x_{-+}(x_{+-})\leq 0.58(1.00) 
\]

From $(x,z)$ plot we obtain following bounds on weak phase $\gamma $ and
strong phases $\delta ^{\prime }s$ $z_{-+}>0,\gamma \geq 70^{\circ
},10^{\circ }\leq \delta _{-+}\leq 40^{\circ },$ $z_{-+}<0,\gamma \geq
65^{\circ },$ $(180-\delta _{-+})$. For $\delta _{+-}$ we get $25^{\circ
}\leq \delta _{+-}\leq 90^{\circ }$ or $(180-\delta _{+-}).$
\end{abstract}

The decays of $B$ mesons to two light mesons which belong to an octet or
nonet representation of $SU\left( 3\right) $ have been extensively studied %
\cite{1}.The various decay modes can be related to each other by rotation in 
$SU\left( 3\right) $ space, using the relations \cite{2}

\begin{eqnarray}
\left[ F_{j}^{i},q_{k}\right] &=&\delta _{k}^{i}q_{j}-\frac{1}{3}\delta
_{j}^{i}q_{k}  \label{3} \\
\left[ F_{j}^{i},q^{k}\right] &=&-\delta _{j}^{k}q^{i}+\frac{1}{3}\delta
_{j}^{i}q^{k}  \label{4}
\end{eqnarray}
where $F_{j}^{i}$ are $SU\left( 3\right) $ generators. In particular, using 
\begin{eqnarray}
\left[ F_{2}^{3},H_{W}(s)\right] &=&H_{W}(d)  \label{5} \\
F_{3}^{2}|\bar{s}\rangle &=&-|\bar{d}\rangle ,\text{ }F_{3}^{2}|d\rangle
=|s\rangle  \nonumber \\
F_{2}^{3}|\bar{d}\rangle &=&-|\bar{s}\rangle  \label{6} \\
F_{2}^{3}|\bar{B}_{d}^{0}\rangle &=&F_{2}^{3}|b\bar{d}\rangle =-|b\bar{s}%
\rangle =-|\bar{B}_{s}^{0}\rangle  \label{7}
\end{eqnarray}
we can relate various decay modes. The $SU\left( 3\right) $ relations for
various decay modes have been studied in refrences \cite{3,4}. In this
paper, we concentrate on the decay modes which we believe have not been
covered although there is bound to be overlap.Suppressing the CKM\ factors
(which can be inserted for each relevant amplitude) we obtain the following
relations for various decay modes for $B\rightarrow PV$ decays: 
\begin{eqnarray}
\left\langle K^{*0}\bar{K}^{0}\left| H_{W}(d)\right| \bar{B}%
_{d}^{0}\right\rangle &=&\left\langle K^{*0}\bar{K}^{0}\left|
H_{W}(s)\right| \bar{B}_{s}^{0}\right\rangle  \nonumber \\
&&+\left\{ 
\begin{array}{c}
\frac{1}{\sqrt{2}}\left[ \left\langle \rho ^{0}\bar{K}^{0}\left| H_{W}\left(
s\right) \right| \bar{B}_{d}^{0}\right\rangle -\left\langle \omega \bar{K}%
^{0}\left| H_{W}\left( s\right) \right| \bar{B}_{d}^{0}\right\rangle \right]
\\ 
-\left\langle \phi \bar{K}^{0}\left| H_{W}\left( s\right) \right| \bar{B}%
_{d}^{0}\right\rangle%
\end{array}
\right\}  \nonumber \\
&=&\left. 
\begin{array}{c}
\left\langle K^{*0}\bar{K}^{0}\left| H_{W}(s)\right| \bar{B}%
_{s}^{0}\right\rangle + \\ 
\{\frac{1}{2}\left[ \left( C^{\prime }-P^{\prime }-P_{EW}^{\prime }\right)
-\left( C^{\prime }+P^{\prime }+\frac{1}{3}P_{EW}^{\prime }\right) \right]
\\ 
-\left( -P^{\prime }+\frac{1}{3}P_{EW}^{\prime }\right) \}%
\end{array}
\right.  \nonumber \\
&&  \label{8} \\
\left\langle K^{0}\bar{K}^{*0}\left| H_{W}\left( d\right) \right| \bar{B}%
_{d}^{0}\right\rangle &=&\left. 
\begin{array}{c}
\left\langle K^{0}\bar{K}^{*0}\left| H_{W}\left( s\right) \right| \bar{B}%
_{d}^{0}\right\rangle \\ 
+\left\{ 
\begin{array}{c}
\frac{1}{\sqrt{2}}\left[ 
\begin{array}{c}
\left\langle \pi ^{0}\bar{K}^{*0}\left| H_{W}\left( s\right) \right| \bar{B}%
_{d}^{0}\right\rangle - \\ 
\left\langle \eta _{ns}\bar{K}^{*0}\left| H_{W}(s)\right| \bar{B}%
_{d}^{0}\right\rangle%
\end{array}
\right] \\ 
+\left\langle \eta _{s}\bar{K}^{*0}\left| H_{W}(s)\right| \bar{B}%
_{d}^{0}\right\rangle%
\end{array}
\right\}%
\end{array}
\right.  \label{9} \\
\text{where \qquad }\eta _{ns} &=&\frac{1}{\sqrt{3}}\left( \eta _{8}+\sqrt{2}%
\eta _{1}\right) ,\text{ }\eta _{s}=\frac{1}{\sqrt{3}}\left( \eta _{1}-\sqrt{%
2}\eta _{8}\right)  \label{10} \\
\left\langle K^{*+}K^{-}\left| H_{W}(d)\right| \bar{B}_{d}^{0}\right\rangle
&=&\left\langle K^{*+}K^{-}\left| H_{W}(s)\right| \bar{B}_{s}^{0}\right%
\rangle -\left\langle \rho ^{+}K^{-}\left| H_{W}(s)\right| \bar{B}%
_{d}^{0}\right\rangle  \label{11} \\
\left\langle K^{*-}K^{+}\left| H_{W}(d)\right| \bar{B}_{d}^{0}\right\rangle
&=&\left\langle K^{+}K^{*-}\left| H_{W}(s)\right| \bar{B}_{s}^{0}\right%
\rangle -\left\langle \pi ^{+}K^{*-}\left| H_{W}(s)\right| \bar{B}%
_{d}^{0}\right\rangle  \label{12} \\
\left\langle K^{*0}K^{-}\left| H_{W}(d)\right| B^{-}\right\rangle &=&\frac{1%
}{\sqrt{2}}\left[ \left\langle \rho ^{0}K^{-}\left| H_{W}(s)\right|
B^{-}\right\rangle -\left\langle \omega K^{-}\left| H_{W}(s)\right|
B^{-}\right\rangle \right]  \nonumber \\
&&-\left\langle \phi K^{-}\left| H_{W}(s)\right| B^{-}\right\rangle 
\nonumber \\
&=&\frac{1}{2}\left[ \left( T^{\prime }+C^{\prime }+P^{\prime
}+P_{EW}^{\prime }\right) -\left( T^{\prime }+C^{\prime }+P^{\prime }+\frac{1%
}{3}P_{EW}^{\prime }\right) \right]  \nonumber \\
&&-\left[ -P^{\prime }+\frac{1}{3}P_{EW}^{\prime }\right]  \label{13}
\end{eqnarray}
\begin{eqnarray}
\left\langle K^{0}K^{*-}\left| H_{W}(d)\right| B^{-}\right\rangle &=&\frac{1%
}{\sqrt{2}}\left[ \left\langle \pi ^{0}K^{*-}\left| H_{W}(s)\right|
B^{-}\right\rangle -\left\langle \eta _{ns}K^{*-}\left| H_{W}(s)\right|
B^{-}\right\rangle \right]  \nonumber \\
&&+\left\langle \eta _{s}K^{*-}\left| H_{W}(s)\right| B^{-}\right\rangle
\label{14}
\end{eqnarray}
For the sake of completeness, using the above technique we derive the
following relations . 
\begin{eqnarray}
\left\langle \rho ^{-}\pi ^{+}\left| H_{W}\left( d\right) \right| \bar{B}%
_{d}^{0}\right\rangle &=&\left\langle K^{*-}\pi ^{+}\left| H_{W}\left(
s\right) \right| \bar{B}_{d}^{0}\right\rangle +\left\langle \rho ^{-}\pi
^{+}\left| H_{W}\left( s\right) \right| \bar{B}_{s}^{0}\right\rangle
\label{15} \\
\left\langle \rho ^{+}\pi ^{-}\left| H_{W}\left( d\right) \right| \bar{B}%
_{d}^{0}\right\rangle &=&\left\langle \rho ^{+}K^{-}\left| H_{W}\left(
s\right) \right| \bar{B}_{d}^{0}\right\rangle +\left\langle \rho ^{+}\pi
^{-}\left| H_{W}\left( s\right) \right| \bar{B}_{s}^{0}\right\rangle
\label{16} \\
\left\langle \phi \pi ^{-}\left| H_{W}\left( d\right) \right|
B^{-}\right\rangle &=&\left\langle \phi K^{-}\left| H_{W}\left( s\right)
\right| B^{-}\right\rangle +\left\langle \bar{K}^{*0}\pi ^{-}\left|
H_{W}\left( s\right) \right| B^{-}\right\rangle  \label{17} \\
\left\langle \phi \pi ^{0}\left| H_{W}\left( d\right) \right| \bar{B}%
_{d}^{0}\right\rangle &=&-\frac{1}{\sqrt{2}} 
\begin{array}{c}
\left\langle \phi \bar{K}^{0}\left| H_{W}\left( s\right) \right| \bar{B}%
_{d}^{0}\right\rangle +\left\langle \bar{K}^{*0}\pi ^{0}\left| H_{W}\left(
s\right) \right| \bar{B}_{d}^{0}\right\rangle \\ 
+\left\langle \phi \pi ^{0}\left| H_{W}\left( s\right) \right| \bar{B}%
_{d}^{0}\right\rangle%
\end{array}
\nonumber \\
&=&-\frac{1}{\sqrt{2}}\left[ -P^{\prime }-\frac{1}{3}P_{EW}^{\prime }\right]
+\frac{1}{\sqrt{2}}\left[ C^{\prime }-P^{\prime }+P_{EW}^{\prime }\right] 
\nonumber \\
&&+\frac{1}{\sqrt{2}}\left[ -C^{\prime }-P_{EW}^{\prime }\right]  \label{18}
\end{eqnarray}
We note that only penguin contributes to the decay channel $\bar{B}%
^{0}\rightarrow K^{*0}\bar{K}^{0},\bar{B}^{0}\rightarrow K^{0}\bar{K}^{*0}.$
Hence it follows from Eqs.(\ref{8}), (\ref{9}), (\ref{13}) and (\ref{14}) 
\begin{eqnarray}
\left\langle K^{*0}\bar{K}^{0}\left| H_{W}\left( d\right) \right| \bar{B}%
_{d}^{0}\right\rangle &=&P=\left| V_{ub}\right| \left| V_{ud}^{*}\right|
e^{-i\gamma }P_{u}-\left| V_{cb}\right| \left| V_{cd}^{*}\right| P_{c} 
\nonumber \\
&=&\left| V_{cb}\right| \left| V_{cs}\right| \left| P_{c}\right| \times 
\left[ -\bar{\lambda}+re^{i\left( -\gamma +\delta \right) }\right]
\label{20} \\
\left\langle K^{*0}\bar{K}^{0}\left| H_{W}\left( s\right) \right| \bar{B}%
_{s}^{0}\right\rangle &=&P^{\prime }=\left| V_{ub}\right| \left|
V_{us}^{*}\right| e^{-i\gamma }P_{u}+\left| V_{cb}\right| \left|
V_{cs}^{*}\right| P_{c}  \nonumber \\
&\approx &\left| V_{cb}\right| \left| V_{cs}\right| \left| P_{c}\right|
\label{20i} \\
\bar{A}\left( \bar{B}_{d}^{0}\rightarrow K^{0}\bar{K}^{*0}\right) &=&\left|
V_{cb}\right| \left| V_{cs}\right| \left| \tilde{P}_{c}\right| \times \left[
-\bar{\lambda}+\tilde{r}e^{i\left( -\gamma +\delta \right) }\right]
\label{20j} \\
\bar{A}\left( \bar{B}_{s}^{0}\rightarrow K^{0}\bar{K}^{*0}\right) &\approx
&\left| V_{cb}\right| \left| V_{cs}\right| \left| \tilde{P}_{c}\right|
\label{20k} \\
\bar{A}\left( B^{-}\rightarrow K^{*0}K^{-}\right) &=&\left| V_{cb}\right|
\left| V_{cs}\right| \left| P_{c}^{\prime 0-}\right| \left[ -\bar{\lambda}%
+r_{0-}^{\prime }e^{i\left( -\gamma +\delta _{0-}^{\prime }\right) }\right]
\label{20a} \\
\bar{A}\left( B^{-}\rightarrow \phi K^{-}\right) &=&\left| V_{cb}\right|
\left| V_{cs}\right| \left| P_{c}^{\prime 0-}\right| \left[ 1+\frac{1}{3}%
\delta _{EW}^{0-}\right]  \label{20b} \\
\bar{A}\left( B^{-}\rightarrow K^{0}K^{*-}\right) &=&\left| V_{cb}\right|
\left| V_{cs}\right| \left| P_{c}^{\prime -0}\right| \left[ -\bar{\lambda}%
+r_{-0}^{\prime }e^{i\left( -\gamma +\delta _{-0}^{\prime }\right) }\right]
\label{20c} \\
\bar{A}\left( B^{-}\rightarrow \eta _{s}K^{*-}\right) &=&\left|
V_{cb}\right| \left| V_{cs}\right| \left| P_{c}^{\prime -0}\right| \left[ 1+%
\frac{1}{3}\delta _{EW}^{-0}\right]  \label{20d} \\
&&  \nonumber
\end{eqnarray}

where 
\begin{equation}
re^{i\delta }=\frac{\left| V_{ub}\right| \left| V_{ud}^{*}\right| }{\left|
V_{cb}\right| \left| V_{cs}\right| }\left| \frac{P_{u}}{P_{c}}\right|
e^{i\delta }  \label{21}
\end{equation}
\begin{eqnarray}
V_{cs} &=&1-\frac{\lambda ^{2}}{2},V_{cd}=-\lambda ,V_{us}=\lambda ,V_{ud}=1-%
\frac{\lambda ^{2}}{2}  \nonumber \\
\bar{\lambda} &=&\frac{\lambda }{1-\lambda ^{2}/2}  \label{22}
\end{eqnarray}
and 
\[
\delta _{EW}=\frac{\left| P_{EW}^{\prime }\right| }{\left| P_{c}^{\prime
}\right| } 
\]

It may also be noted, that for the decay channel $B^{-}\rightarrow \phi \pi
^{-},$ only color supressed and electroweak penguin contributes. Negelecting
the color supressed penguin contribution in Eq.(\ref{17})$,$ we get 
\begin{equation}
\bar{A}\left( B^{-}\rightarrow \bar{K}^{*0}\pi ^{-}\right) =\left|
V_{cb}\right| \left| V_{cs}\right| \left| P_{c}^{\prime 0-}\right|
\label{33}
\end{equation}
and 
\begin{equation}
\bar{A}\left( B^{-}\rightarrow \rho ^{-}\bar{K}^{0}\right) =\left|
V_{cb}\right| \left| V_{cs}\right| \left| P_{c}^{\prime -0}\right|
\label{35}
\end{equation}
Electroweak penguins are in a class different from those of gluon penguins.
Assumung factorization for electroweak penguin, we get from Eq.(\ref{8})$,$
an intresting sum rule 
\begin{equation}
f_{\rho }F_{1}^{B-K}\left( m_{\rho }^{2}\right) -\frac{1}{3}f_{\omega
}F_{1}^{B-K}\left( m_{\omega }^{2}\right) -\frac{2}{3}f_{\phi
}F_{1}^{B-K}\left( m_{\phi }^{2}\right) =0  \label{36}
\end{equation}
Assuming 
\begin{equation}
F_{1}^{B-K}\left( m_{\rho }^{2}\right) =F_{1}^{B-K}\left( m_{\omega
}^{2}\right) =F_{1}^{B-K}\left( m_{\phi }^{2}\right)  \label{37}
\end{equation}
we obtain the relation 
\begin{equation}
f_{\rho }-\frac{1}{3}f_{\omega }-\frac{2}{3}f_{\phi }=0  \label{38}
\end{equation}
reminiscent of current algebra, and spectral function sum rules of 1960's %
\cite{5}. Sum rule (\ref{38}) is very well satisfied by the experimental
values \cite{6} 
\[
f_{\rho }=\left( 209\pm 1\right) MeV,\text{ }f_{\omega }=187\pm 3MeV,\text{ }%
f_{\phi }=221\pm 3MeV 
\]
The assumption stated in Eq (\ref{37}) holds as the form factor $F_{1}$ at
1GeV will not differ much at the masses $m_{\rho }^{2},m_{\omega }^{2}$ and $%
m_{\phi }^{2}$

Similarly we get the sum rule 
\begin{equation}
f_{\pi }-\frac{1}{3}f_{\eta _{ns}}-\frac{2}{3}f_{\eta _{s}}=0  \label{39}
\end{equation}
Finally, from the term in curly bracket in Eq.(\ref{8}) we get the following
relations between average decay rates and mixing induced $CP$ asymmetries
for $B_{d}^{0}\rightarrow \rho ^{0}\bar{K}^{0}$,$\omega \bar{K}^{0}$, $\phi 
\bar{K}^{0}$%
\begin{eqnarray}
\frac{\left\langle \Gamma \right\rangle _{\omega K}+\left\langle \Gamma
\right\rangle _{\rho ^{0}K}}{\left\langle \Gamma \right\rangle _{\phi K}}
&\approx &1\left[ \frac{\left( 4.7\pm 0.6\right) +\left( 5.1\pm 1.6\right) }{%
8.3\pm 1.12}=1.2\pm 0.3\right] _{\text{expt. value}}  \label{39a} \\
\frac{S(\rho ^{0}K_{s})+S(\omega K_{s})}{2} &=&S(\phi K_{s})=-\sin 2\beta
\label{39b} \\
C(\rho ^{0}K_{s}) &=&-C(\omega K_{s})\approx 2r_{c}^{\prime }\sin \gamma
\sin \delta ^{\prime }+O(r_{c}^{\prime }r_{EW}^{\prime })  \label{39c}
\end{eqnarray}
where 
\begin{equation}
r_{c}^{\prime }e^{i\delta ^{\prime }}=\frac{C^{\prime }}{P^{\prime }}=\frac{%
\left| C^{\prime }\right| }{\left| P^{\prime }\right| }e^{i\delta ^{\prime }}
\label{39d}
\end{equation}
We obtain some intresting results from Eqs.(\ref{11}) and (\ref{12})$.$
First we note that only $W$-exchange diagram contributes to $\bar{B}%
_{d}^{0}\rightarrow K^{*+}K^{-},$ $K^{+}K^{*-}$ decay channels; neglecting
this contribution and also neglecting the $W$-exchange contribution to $\bar{%
B}_{s}^{0}\rightarrow \rho ^{+}\pi ^{-},\rho ^{-}\pi ^{+}$ in Eqs (\ref{15})
and (\ref{16})$,$ we get 
\begin{eqnarray}
\left\langle K^{*+}K^{-}\left| H_{W}\left( s\right) \right| \bar{B}%
_{s}^{0}\right\rangle &=&\left\langle \rho ^{+}K^{-}\left| H_{W}\left(
s\right) \right| \bar{B}_{d}^{0}\right\rangle  \nonumber \\
&=&\left| V_{ub}\right| \left| V_{ud}^{*}\right| \left| T^{+-}\right| \left[
e^{-i\gamma }\bar{\lambda}\frac{f_{K}}{f_{\pi }}-\frac{1}{\bar{\lambda}}%
r_{+-}e^{i\delta _{+-}}\right]  \label{40} \\
r_{+-} &=&\frac{\left| P^{+-}\right| }{\left| T^{+-}\right| }  \nonumber \\
\left\langle K^{+}K^{*-}\left| H_{W}\left( s\right) \right| \bar{B}%
_{s}^{0}\right\rangle &=&\left\langle \pi ^{+}K^{*-}\left| H_{W}\left(
s\right) \right| \bar{B}_{d}^{0}\right\rangle  \nonumber \\
&=&\left| V_{ub}\right| \left| V_{ud}^{*}\right| \left| T^{-+}\right| \left[
e^{-i\gamma }\bar{\lambda}\frac{f_{K^{*}}}{f_{\rho }}-\frac{1}{\bar{\lambda}}%
r_{-+}e^{i\delta _{-+}}\right]  \label{41}
\end{eqnarray}
Before, we examine the consequences of $B\rightarrow PV$ decays for various
channels obtained above, we discuss the various decay channels of $%
B\rightarrow P_{1}P_{2}$. For these decays, replace $K^{*}$ with $K$ in Eqs.$%
($\ref{9}$),($\ref{12}$)$ and $($\ref{14}$).$Hence for the decay amplitudes $%
\bar{A}(\bar{B}_{d}^{0}\rightarrow K^{0}\bar{K}^{0})$, $A(\bar{B}%
_{s}^{0}\rightarrow K^{0}\bar{K}^{0})$; replace $P_{C}$ by $p_{c}^{\prime }$%
; $r,\delta $ by $r_{n},\delta _{n}^{\prime }$ in Eqs. (\ref{20}) and (\ref%
{20i}) and $P_{c}^{\prime 0-},r_{0-}^{\prime }$ and $\delta _{0-}^{\prime }$
by $p_{c}^{\prime },r_{c}^{\prime }$ and $\delta _{c}^{\prime }$ in Eq.(\ref%
{20a}) for the amplitude $\bar{A}(B^{-}\rightarrow K^{0}K^{-}).$ The
amplitudes for the decays $\bar{B}_{s}^{0}\rightarrow K^{+}K^{-},$ $%
B^{-}\rightarrow \eta _{s}K^{-}$ and $B^{-}\rightarrow \pi ^{-}\bar{K}^{0}$
are given below.

\begin{eqnarray}
\bar{A}\left( \bar{B}_{s}^{0}\rightarrow K^{+}K^{-}\right) &=&\bar{A}\left( 
\bar{B}_{d}^{0}\rightarrow \pi ^{+}K^{-}\right)  \nonumber \\
&=&\left| V_{cb}\right| \left| V_{us}\right| \left| t^{\prime }\right|
e^{-i\gamma }+\left| V_{cb}\right| \left| V_{us}\right| \left| p_{c}^{\prime
}\right|  \nonumber \\
&=&\left| V_{cb}\right| \left| V_{us}\right| \left| p_{c}^{\prime }\right| 
\left[ 1+r^{\prime }e^{i\left( -\gamma +\delta ^{\prime }\right) }\right]
\label{47b} \\
\bar{A}\left( B^{-}\rightarrow K^{0}K^{-}\right) &=&\left| V_{cb}\right|
\left| V_{cs}\right| \left| p_{c}^{\prime }\right| \left[ -\bar{\lambda}%
+r_{c}^{\prime }e^{i\left( -\gamma +\delta _{c}^{\prime }\right) }\right]
\label{48} \\
\bar{A}\left( B^{-}\rightarrow \eta _{s}K^{-}\right) &=&\left| V_{cb}\right|
\left| V_{cs}\right| \left[ p_{c}^{\prime }+\frac{1}{3}p_{EW}^{\prime }%
\right]  \nonumber  \label{49}
\end{eqnarray}
In addition we have 
\begin{equation}
\bar{A}\left( B^{-}\rightarrow \pi ^{-}\bar{K}^{0}\right) \approx \left|
V_{cb}\right| \left| V_{cs}\right| \left| p_{c}^{\prime }\right|  \label{48b}
\end{equation}
For the observables for the decay $B\rightarrow P_{1}P_{2}$: 
\begin{eqnarray}
\left\langle \Gamma _{n}\right\rangle ,\Delta _{n} &=&\frac{\Gamma \left(
B_{d}^{0}\rightarrow K^{0}\bar{K}^{0}\right) \pm \Gamma \left( \bar{B}%
_{d}^{0}\rightarrow K^{0}\bar{K}^{0}\right) }{2}  \label{51} \\
\left\langle \Gamma _{c}\right\rangle ,\Delta _{c} &=&\frac{\Gamma \left(
B^{+}\rightarrow K^{+}\bar{K}^{0}\right) \pm \Gamma \left( B^{-}\rightarrow
K^{-}K^{0}\right) }{2}  \label{52} \\
A_{CP}^{n} &=&\frac{\Delta _{n}}{\left\langle \Gamma _{n}\right\rangle }
\label{53} \\
A_{CP}^{c} &=&\frac{\Delta _{c}}{\left\langle \Gamma _{c}\right\rangle }
\label{54}
\end{eqnarray}
we obtain the following results 
\begin{eqnarray}
\Gamma \left( \bar{B}_{s}^{0}\rightarrow K^{0}\bar{K}^{0}\right) &\approx
&\Gamma \left( B^{-}\rightarrow \pi ^{-}\bar{K}^{0}\right)  \label{55} \\
\frac{\left\langle \Gamma _{n}\right\rangle }{\bar{\lambda}^{2}\Gamma \left(
B^{-}\rightarrow \pi ^{-}\bar{K}^{0}\right) } &\approx &\left[ 1-\frac{%
2r_{n}^{\prime }}{\bar{\lambda}}\cos \gamma \cos \delta _{n}^{\prime }+\frac{%
r_{n}^{\prime 2}}{\bar{\lambda}^{2}}\right]  \label{56} \\
\frac{\left\langle \Gamma _{c}\right\rangle }{\bar{\lambda}^{2}\Gamma \left(
B^{-}\rightarrow \pi ^{-}\bar{K}^{0}\right) } &\approx &\left[ 1-\frac{%
2r_{c}^{\prime }}{\bar{\lambda}}\cos \gamma \cos \delta _{n}^{\prime }+\frac{%
r_{c}^{\prime 2}}{\bar{\lambda}^{2}}\right]  \label{57} \\
A_{CP}^{n,c} &\approx &\frac{\bar{\lambda}^{2}\Gamma \left( B^{-}\rightarrow
\pi ^{-}\bar{K}^{0}\right) }{\left\langle \Gamma _{n,c}\right\rangle }\left[
2\frac{r_{n,c}^{\prime }}{\bar{\lambda}}\sin \delta _{n,c}^{\prime }\sin
\gamma \right]  \label{57a}
\end{eqnarray}
Using the following values for the branching ratios: \cite{8} 
\begin{eqnarray}
\mathcal{B}\left( B^{-}\rightarrow \pi ^{-}\bar{K}^{0}\right) &=&\left(
24.1\pm 1.7\right) \times 10^{-6}  \label{58} \\
\mathcal{B}\left( B^{-}\rightarrow K^{0}K^{-}\right) &=&\left( 1.2\pm
0.32\right) \times 10^{-6}  \nonumber \\
&&  \label{59} \\
\mathcal{B}\left( B^{0}\rightarrow K^{0}\bar{K}^{0}\right) &=&\left( 1.13\pm
0.38\right) \times 10^{-6}  \nonumber
\end{eqnarray}
we get 
\begin{eqnarray}
\frac{\left\langle \Gamma _{n}\right\rangle }{\bar{\lambda}^{2}\Gamma \left(
B^{-}\rightarrow \pi ^{-}\bar{K}^{0}\right) } &=&0.94\pm 0.32  \label{60} \\
\frac{\left\langle \Gamma _{c}\right\rangle }{\bar{\lambda}^{2}\Gamma \left(
B^{-}\rightarrow \pi ^{-}\bar{K}^{0}\right) } &=&1.00\pm 0.25  \label{61}
\end{eqnarray}
Hence 
\begin{eqnarray}
\left( 1-\frac{2r_{n}^{\prime }}{\bar{\lambda}}\cos \gamma \cos \delta
_{n}^{\prime }+\frac{r_{n}^{\prime 2}}{\bar{\lambda}^{2}}\right) &=&0.94\pm
0.32  \label{61a} \\
\left( 1-\frac{2r_{c}^{\prime }}{\bar{\lambda}}\cos \gamma \cos \delta
_{c}^{\prime }+\frac{r_{c}^{\prime 2}}{\bar{\lambda}^{2}}\right) &=&1.00\pm
0.25  \label{61b}
\end{eqnarray}
In this approximation 
\begin{equation}
A_{CP}^{n,c}=\frac{2r_{n,c}^{\prime }}{\bar{\lambda}}\sin \delta
_{n,c}^{\prime }\sin \gamma  \label{63}
\end{equation}
With improvement in experimental accuracy, these predictions can be tested .

From Eq.(\ref{49}), we get the results 
\begin{eqnarray}
\Gamma \left( \bar{B}_{s}^{0}\rightarrow K^{+}K^{-}\right) &\approx &\Gamma
\left( \bar{B}_{d}^{0}\rightarrow \pi ^{+}K^{-}\right)  \label{64} \\
A_{CP}\left( K^{+}K^{-}\right) &\approx &A_{CP}\left( \pi ^{+}K^{-}\right) =%
\frac{-2r^{\prime }\sin \delta _{c}^{\prime }\sin \gamma }{1+2r^{\prime
}\cos \delta _{c}^{\prime }\cos \gamma +r^{\prime 2}}  \nonumber \\
&&  \label{65}
\end{eqnarray}
The experimental values \cite{8} for the decay rates and $A_{CP}$ are 
\begin{eqnarray*}
\Gamma \left( \bar{B}_{d}^{0}\rightarrow \pi ^{+}K^{-}\right) &=&\left(
1.82\pm 0.08\right) \times 10^{-5} \\
\Gamma \left( \bar{B}_{s}^{0}\rightarrow K^{+}K^{-}\right) &<&5.9\times
10^{-5} \\
A_{CP}(\pi ^{+}K^{-}) &=&-0.113\pm 0.020
\end{eqnarray*}

Eqs.(\ref{64}) and (\ref{65}) can be tested when the experimental data on $%
\bar{B}_{s}^{0}\rightarrow K^{+}K^{-}$ will be available.

In order to discuss the mixing induced $CP$-asymmetries we first give a
general expression for the time dependent decay rates for $B^{0}\rightarrow
f,\bar{f}$ in terms of these asymmetries 
\begin{equation}
\left. \Gamma _{\bar{f},f}\left( t\right) =e^{-\Gamma t}\frac{1}{2}\left(
R_{f}+R_{\bar{f}}\right) \left( 1\pm A_{CP}\right) \left\{ 
\begin{array}{c}
1+\cos \Delta mt\left( C_{f}\pm \Delta C_{f}\right) \\ 
-\sin \Delta mt\left( S_{f}\pm \Delta S_{f}\right)%
\end{array}
\right\} \right.  \label{66}
\end{equation}
The decay rates $\bar{\Gamma}_{\bar{f},f}$ can be obtained from Eq.(\ref{66}%
) by changing $\cos \Delta mt\rightarrow -\cos \Delta mt$, $\sin \Delta
mt\rightarrow -\sin \Delta mt.$

The direct $CP$ asymmetries $C_{f}\pm \Delta C_{f}$ and mixing induced $CP$%
-asymmetries $S_{f}\pm \Delta S_{f}$ in terms of scattering amplitudes are
given by 
\begin{eqnarray}
C_{f}\pm \Delta C_{f} &=&\frac{\left| A_{\bar{f},f}\right| ^{2}-\left| 
\bar{A}_{\bar{f},f}\right| ^{2}}{\left| A_{\bar{f},f}\right| ^{2}+\left| 
\bar{A}_{\bar{f},f}\right| ^{2}}  \nonumber \\
&=&\frac{\pm \left( R_{f}-R_{\bar{f}}\right) +\left( R_{f}a_{f}+R_{\bar{f}%
}a_{\bar{f}}\right) }{\left( R_{f}+R_{\bar{f}}\right) \left( 1\pm
A_{CP}\right) }  \label{67} \\
\left( R_{f}+R_{\bar{f}}\right) \left( 1\pm A_{CP}\right) \left( S_{f}\pm
\Delta S_{f}\right) &=&2\func{Im}\left[ e^{-2i\phi _{M}}A_{\bar{f},f}^{*}%
\bar{A}_{\bar{f},f}\right]  \label{68} \\
R_{f} &=&\frac{\left| A_{\bar{f}}\right| ^{2}+\left| \bar{A}_{f}\right| ^{2}%
}{2},\text{ }R_{\bar{f}}=\frac{\left| A_{f}\right| ^{2}+\left| \bar{A}_{\bar{%
f}}\right| ^{2}}{2}  \label{69} \\
&&\left. 
\begin{array}{c}
R_{f}a_{f}=\frac{1}{2}\left[ \left| A_{\bar{f}}\right| ^{2}-\left| \bar{A}%
_{f}\right| ^{2}\right] , \\ 
\text{ }R_{\bar{f}}a_{\bar{f}}=\frac{1}{2}\left[ \left| A_{f}\right|
^{2}-\left| \bar{A}_{\bar{f}}\right| ^{2}\right]%
\end{array}
\right.  \label{69a} \\
A_{CP} &=&\frac{R_{f}a_{f}-R_{\bar{f}}a_{\bar{f}}}{R_{f}+R_{\bar{f}}}
\label{69b} \\
A_{CP}^{\bar{f}} &=&-a_{f},\text{ }A_{CP}^{f}=-a_{\bar{f}}  \label{70}
\end{eqnarray}
Note the weak phase $\phi _{M}=\beta $ for $B_{d}^{0}$ and $\phi _{M}=0$ for 
$B_{s}^{0}.$ For the case $|\bar{f}\rangle =CP|f\rangle ,$ we have $R_{f}=R_{%
\bar{f}},$ $a_{\bar{f}}=a_{f},$ $A_{CP}=0,\Delta C_{f}=0,\Delta S_{f}=0$

For this case, we have 
\begin{eqnarray}
\frac{\Gamma _{f}\left( t\right) -\bar{\Gamma}_{f}\left( t\right) }{\Gamma
_{f}\left( t\right) +\bar{\Gamma}_{f}\left( t\right) } &=&\cos \Delta
mtC_{f}-\sin \Delta mtS_{f}  \label{71} \\
C_{f} &=&\frac{\left| A_{f}\right| ^{2}-\left| \bar{A}_{f}\right| ^{2}}{%
\left| A_{f}\right| ^{2}+\left| \bar{A}_{f}\right| ^{2}}  \label{72} \\
S_{f} &=&\frac{2\func{Im}\left[ e^{-2i\phi _{M}}A_{f}^{*}\bar{A}_{f}\right] 
}{\left| A_{f}\right| ^{2}+\left| \bar{A}_{f}\right| ^{2}}  \label{73}
\end{eqnarray}
The mixing-induced $CP$ asymmetry $S\left( \pi ^{+}K^{-}\right) $ is zero,
since the decay $\bar{B}_{d}^{0}\rightarrow \pi ^{-}K^{+}$ is not allowed in
the standard Model. However, for the decay $\bar{B}_{d}^{0}\rightarrow K^{0}%
\bar{K}^{0}$ and $\bar{B}_{s}^{0}\rightarrow K^{+}K^{-},$ the $CP$
asymmetries, can be easily calculated using Eqs. (\ref{71},\ref{72},\ref{73}%
). We get 
\begin{eqnarray}
C_{n} &=&A_{CP}^{n}  \label{74} \\
S_{n} &=&-\frac{\bar{\lambda}^{2}\Gamma \left( B^{-}\rightarrow \pi ^{-}\bar{%
K}^{0}\right) }{\left\langle \Gamma _{n}\right\rangle }\left\{ 
\begin{array}{c}
\sin 2\beta -2\frac{r_{n}^{\prime }}{\bar{\lambda}}\sin \left( \gamma
+2\beta \right) \cos \delta _{n}^{\prime } \\ 
+\frac{r_{n}^{\prime 2}}{\bar{\lambda}^{2}}\sin \left( 2\beta +2\gamma
\right)%
\end{array}
\right\}  \label{74a}
\end{eqnarray}
\begin{eqnarray}
S_{n}\left( K^{+}K^{-}\right) &=&\frac{-\sin 2\gamma -2r^{\prime }\cos
\delta _{c}^{\prime }\sin \gamma }{1+2r^{\prime }\cos \delta _{c}^{\prime
}\cos \gamma +r^{\prime 2}}  \nonumber \\
&\approx &-\sin 2\gamma +2r^{\prime }\cos \delta _{c}^{\prime }\sin \gamma
\cos 2\gamma  \label{75}
\end{eqnarray}
The mixing induced $CP$-asymmetry $S\left( K^{+}K^{-}\right) $ and direct $%
CP $-asymmetry $A_{CP}\left( \pi ^{+}K^{-}\right) $ would give an alternate
way of determining phase $\gamma ,$ when experimental data on $\bar{B}%
_{s}^{0}\rightarrow K^{+}K^{-}$ will become available.

For the decay $B^{0}\rightarrow PV,$ we get similar results as for $%
B^{0}\rightarrow P_{1}P_{2}$. In particular, the following results are of
particular interest. From Eq\.{s}. (\ref{36}) to (\ref{39}), we obtain 
\begin{eqnarray}
\frac{\left\langle \Gamma ^{0-}\right\rangle }{\bar{\lambda}^{2}\Gamma
\left( B^{-}\rightarrow \bar{K}^{*0}\pi ^{-}\right) } &=&\left[ 1-2\frac{%
r_{0-}^{\prime }}{\bar{\lambda}}\cos \gamma \cos \delta _{0-}^{\prime }+%
\frac{r_{0-}^{2}}{\bar{\lambda}^{2}}\right]  \label{76} \\
\frac{\left\langle \Gamma ^{-0}\right\rangle }{\bar{\lambda}^{2}\Gamma
\left( B^{-}\rightarrow \rho ^{-}\bar{K}^{0}\right) } &=&\left[ 1-2\frac{%
r_{-0}^{\prime }}{\bar{\lambda}}\cos \gamma \cos \delta _{0-}^{\prime }+%
\frac{r_{-0}^{2}}{\bar{\lambda}^{2}}\right]  \label{77} \\
a_{CP}^{0-} &=&\frac{\bar{\lambda}^{2}\Gamma \left( B^{-}\rightarrow \bar{K}%
^{*0}\pi ^{-}\right) }{\left\langle \Gamma ^{0-}\right\rangle }\left[ 2\frac{%
r_{0-}^{\prime }}{\bar{\lambda}}\sin \gamma \sin \delta _{0-}^{\prime }%
\right]  \label{78} \\
a_{CP}^{-0} &=&\frac{\bar{\lambda}^{2}\Gamma \left( B^{-}\rightarrow \rho
^{-}\bar{K}^{0}\right) }{\left\langle \Gamma ^{-0}\right\rangle }\left[ 2%
\frac{r_{-0}^{\prime }}{\bar{\lambda}}\sin \gamma \sin \delta _{-0}^{\prime }%
\right]  \label{79}
\end{eqnarray}
There is no experimental data on $B^{-}\rightarrow K^{*0}K^{-}$ and $%
B^{-}\rightarrow K^{*-}\bar{K}^{0}$ decays to draw conclusion regarding weak
phase $\gamma $ and strong phases $\delta _{0-}^{\prime }$ and $\delta
_{-0}^{\prime }.$ However from Eqs.(\ref{40}) and (\ref{41}), we obtain
observables for which experimental data is available. In particular, we get 
\begin{eqnarray}
\Gamma \left( \bar{B}_{s}^{0}\rightarrow K^{*-}K^{+}\right) &=&\Gamma \left( 
\bar{B}^{0}\rightarrow K^{*-}\pi ^{+}\right)  \label{80} \\
\Gamma \left( \bar{B}_{s}^{0}\rightarrow K^{*+}K^{-}\right) &=&\Gamma \left( 
\bar{B}^{0}\rightarrow \rho ^{+}K^{-}\right)  \label{81} \\
A_{CP}^{\prime \pm } &=&-a^{\prime }\left( K^{*-}K^{+}\right) =-a^{\prime
}\left( K^{*-}\pi ^{+}\right) =\frac{\left| T^{-+}\right| ^{2}}{%
R_{-+}^{\prime }}\left[ 2r_{-+}\frac{f_{K^{*}}}{f_{\rho }}\sin \gamma \sin
\delta _{-+}\right]  \nonumber \\
&&  \label{82} \\
A_{CP}^{\prime \mp } &=&-a\left( K^{*+}K^{-}\right) =-a^{\prime }\left( \rho
^{+}K^{-}\right) =\frac{\left| T^{+-}\right| ^{2}}{R_{+-}^{\prime }}\left[
2r_{+-}\frac{f_{K}}{f_{\pi }}\sin \gamma \sin \delta _{+-}\right]  \label{83}
\\
A_{CP}^{\pm } &=&-a_{-+}=\frac{\left| T^{-+}\right| ^{2}}{R_{-+}}\left[
-2r_{-+}\sin \gamma \sin \delta _{-+}\right]  \label{84} \\
A_{CP}^{\mp } &=&-a_{+-}=\frac{\left| T^{+-}\right| ^{2}}{R_{+-}}\left[
-2r_{+-}\sin \gamma \sin \delta _{+-}\right]  \label{85}
\end{eqnarray}
where 
\begin{eqnarray}
\frac{R_{-+}^{\prime }}{\left| T^{-+}\right| ^{2}} &=&B_{-+}^{\prime }=\left[
\left( \bar{\lambda}\frac{f_{K^{*}}}{f_{\rho }}\right) ^{2}-2r_{-+}\frac{%
f_{K^{*}}}{f_{\rho }}\cos \gamma \cos \delta _{-+}+\frac{1}{\bar{\lambda}^{2}%
}r_{-+}^{2}\right]  \label{86} \\
\frac{R_{+-}^{\prime }}{\left| T^{+-}\right| ^{2}} &=&B_{+-}^{\prime }=\left[
\left( \bar{\lambda}\frac{f_{K}}{f_{\pi }}\right) ^{2}-2r_{+-}\frac{f_{K}}{%
f_{\pi }}\cos \gamma \cos \delta _{+-}+\frac{1}{\bar{\lambda}^{2}}r_{+-}^{2}%
\right]  \label{87} \\
\frac{R_{-+}}{\left| T^{-+}\right| ^{2}} &=&B_{-+}=\left[ 1+2r_{-+}\cos
\gamma \cos \delta _{-+}+r_{-+}^{2}\right]  \label{88} \\
\frac{R_{+-}}{\left| T^{+-}\right| ^{2}} &=&B_{+-}=\left[ 1+2r_{+-}\cos
\gamma \cos \delta _{+-}+r_{+-}^{2}\right]  \label{89}
\end{eqnarray}
From above equations, we obtain the results 
\begin{eqnarray}
\frac{a_{-+}^{\prime }}{a_{-+}} &=&-\left( \frac{f_{K^{*}}}{f_{\rho }}%
\right) \left( \frac{R_{-+}}{R_{-+}^{\prime }}\right)  \label{90} \\
\frac{a_{+-}^{\prime }}{a_{+-}} &=&-\left( \frac{f_{K}}{f_{\pi }}\right)
\left( \frac{R_{+-}}{R_{+-}^{\prime }}\right)  \label{91} \\
r_{-+}^{2} &=&\bar{\lambda}^{2}\left[ B_{-+}\frac{R_{-+}^{\prime }}{R_{-+}}%
+\left( B_{-+}-1\right) \frac{f_{K^{*}}}{f_{\rho }}\right]  \label{92} \\
r_{+-}^{2} &=&\bar{\lambda}^{2}\left[ B_{+-}\frac{R_{+-}^{\prime }}{R_{+-}}%
+\left( B_{+-}-1\right) \frac{f_{K}}{f_{\pi }}\right]  \label{93}
\end{eqnarray}
In the end, we discuss the relations between various observables defined in
Eqs.(\ref{66}-\ref{71}) for the decays $\bar{B}_{d}^{0}\rightarrow \rho
^{-}\pi ^{+},\rho ^{+}\pi ^{-}$ and $\bar{B}_{s}^{0}\rightarrow K^{*-}K^{+},$
$K^{*+}K^{-}.$ The $CP$ violating asymmetry $A_{CP},$ the direct $CP$
violation $C$, the dilution parameter $\Delta C$, the mixing induced $CP\,$%
violation $S$ and dilution parameter $\Delta S$ for the $\rho \pi $ and $%
K^{*}K$ decay channels, are related to each other by $SU(3).$ We obtain the
following results for the various observables, using Eqs. (\ref{40}) and (%
\ref{41}); dashed quantites refer to $\bar{B}_{s}^{0}\rightarrow
K^{*-}K^{+}, $ $K^{*+}K^{-}$ and undashed to $\bar{B}_{d}^{0}\rightarrow
\rho ^{-}\pi ^{+},\rho ^{+}\pi ^{-})$%
\begin{eqnarray}
A_{CP}^{\prime } &=&\left. -\frac{1}{2}\frac{R_{-+}+R_{+-}}{R_{-+}^{\prime
}+R_{+-}^{\prime }}\left[ 
\begin{array}{c}
\left( \frac{f_{K^{*}}}{f_{\rho }}+\frac{f_{K}}{f_{\pi }}\right) A_{CP} \\ 
+\left( \frac{f_{K^{*}}}{f_{\rho }}-\frac{f_{K}}{f_{\pi }}\right) \left(
C+A_{CP}\Delta C\right)%
\end{array}
\right] \right.  \label{93a} \\
C^{\prime }+A_{CP}^{\prime }\Delta C^{\prime } &=&\left. -\frac{1}{2}\frac{%
R_{-+}+R_{+-}}{R_{-+}^{\prime }+R_{+-}^{\prime }}\left[ 
\begin{array}{c}
\left( \frac{f_{K^{*}}}{f_{\rho }}+\frac{f_{K}}{f_{\pi }}\right) \left(
C+A_{CP}\Delta C\right) \\ 
+\left( \frac{f_{K^{*}}}{f_{\rho }}-\frac{f_{K}}{f_{\pi }}\right) A_{CP}%
\end{array}
\right] \right.  \label{93b} \\
\Delta C+A_{CP}C &=&\frac{R_{-+}-R_{+-}}{R_{-+}+R_{+-}};\Delta C^{\prime
}+A_{CP}^{\prime }C^{\prime }=\frac{R_{-+}^{\prime }-R_{+-}^{\prime }}{%
R_{-+}^{\prime }+R_{+-}^{\prime }}  \label{93c}
\end{eqnarray}
Taking $\delta _{t}=\left( \delta _{T}^{+-}-\delta _{T}^{-+}\right) =0$\cite%
{9}, we get for mixing induced $CP$- violating parameters $S$ and $\Delta S$%
, the following results: (neglecting terms of the order $r_{-+}r_{+-}$) 
\begin{eqnarray}
\frac{1}{t}\left( B_{-+}+t^{2}B_{+-}\right) \left[ \Delta S^{\prime
}+A_{CP}^{\prime }S^{\prime }\right] &\approx &\left. \frac{R_{-+}+R_{+-}}{%
R_{-+}^{\prime }+R_{+-}^{\prime }}2\cos \gamma \left[ 
\begin{array}{c}
-\frac{f_{K^{*}}}{f_{\rho }}r_{+-}\sin \delta _{+-} \\ 
+\frac{f_{K}}{f_{\pi }}r_{-+}\cos \delta _{-+}%
\end{array}
\right] \right.  \nonumber \\
&&  \label{93d}
\end{eqnarray}
\begin{eqnarray}
\frac{1}{t}\left( B_{-+}+t^{2}B_{+-}\right) \left[ S+A_{CP}\Delta S\right]
&\approx &\left. 2\left[ 
\begin{array}{c}
-\sin \left( 2\beta +2\gamma \right) -\sin \left( 2\beta +\gamma \right)
\times \\ 
\left( r_{-+}\cos \delta _{-+}+r_{+-}\cos \delta _{+-}\right)%
\end{array}
\right] \right.  \nonumber \\
&&  \nonumber \\
&&  \label{94} \\
\frac{1}{t}\left( B_{-+}+t^{2}B_{+-}\right) \left[ \Delta S+A_{CP}S\right]
&\approx &2\cos \left( 2\beta +\gamma \right) \left[ r_{+-}\sin \delta
_{+-}-r_{-+}\sin \delta _{-+}\right]  \label{94a} \\
\frac{1}{t}\left( B_{-+}+t^{2}B_{+-}\right) \left[ S^{\prime
}+A_{CP}^{\prime }\Delta S^{\prime }\right] &\approx &\frac{R_{-+}+R_{+-}}{%
R_{-+}^{\prime }+R_{+-}^{\prime }}2\left[ 
\begin{array}{c}
\frac{-\bar{\lambda}^{2}f_{K^{*}}f_{K}}{f_{\rho }f_{\pi }}\sin 2\gamma \\ 
+\left. \sin \gamma \left( 
\begin{array}{c}
\frac{f_{K}}{f_{\pi }}r_{-+}\cos \delta _{-+} \\ 
+\frac{f_{K^{*}}}{f_{\rho }}r_{+-}\cos \delta _{+-}%
\end{array}
\right) \right.%
\end{array}
\right]  \nonumber \\
&&  \label{94b}
\end{eqnarray}
In order to make contact with the experimental data, we use the data of
refrence \cite{8}.Using the following experimental values 
\begin{eqnarray}
R_{-+}+R_{+-} &=&\left( 22.8\pm 2.5\right) \times 10^{-6}  \nonumber \\
R_{-+}^{\prime } &=&\left( 11.8\pm 1.5\right) \times 10^{-6},R_{+-}^{\prime
}=\left( 8.5\pm 2.8\right) \times 10^{-6}  \label{96} \\
a_{\bar{-}+} &=&0.15\pm 0.08,\text{ }a_{+-}=0.53\pm 0.13  \nonumber \\
\text{ }C &=&0.30\pm 0.13,\text{ }\Delta C=0.33\pm 0.13  \label{97}
\end{eqnarray}
we obtain 
\begin{equation}
A_{CP}=-0.087\pm 0.07,\,R_{-+}=\left( 14.9\pm 2.2\right) \times
10^{-6},R_{+-}=\left( 7.9\pm 1.8\right) \times 10^{-6}  \label{97a}
\end{equation}

Hence using the above values we get from Eqs.(\ref{90}), (\ref{91}) 
\[
a_{-+}^{\prime }=-0.20\pm 0.11\left( -0.05\pm 0.14\right) \text{, }%
a_{+-}^{\prime }=-0.60\pm 0.41\left( -0.26\pm 0.15\right) 
\]
where the numbers within the brackets are experimental values.

In order to obtain the values of $r_{-+}$ and $r_{+-}$ from Eqs.(\ref{92})
and (\ref{93}) we need the values for $B_{-+}$ and $B_{+-}$. Using \cite{9}, 
\begin{equation}
B_{-+}=1.03\pm 0.12,\text{ }B_{+-}=1.05\pm 0.12  \label{99}
\end{equation}

we obtain 
\begin{eqnarray}
r_{-+}^{2} &=&0.044\pm 0.014;\text{ }r_{-+}=0.21\pm 0.04;\text{ }0.17\leq
r_{-+}\leq 0.24  \label{100} \\
r_{+-}^{2} &=&0.063\pm 0.038,\text{ }r_{+-}=0.25\pm 0.06;\text{ }0.19\leq
r_{+-}\leq 0.30  \label{101}
\end{eqnarray}
to be compared with the values given in refrence \cite{7}. The following
remarks are in order. The values given in Eq.(\ref{99}) has been obtained
from $B_{-+}=\frac{R_{-+}}{\left| T_{-+}\right| ^{2}},B_{+-}=\frac{R_{+-}}{%
\left| T_{+-}\right| ^{2}},$ using factorization for the tree amplitude $%
T_{-+}$ with $\left| V_{ub}\right| =3.45\times 10^{-3},$ $f_{+}(m_{\rho
}^{2})=0.26$ and $t^{2}=0.52$

Using the above values of $r_{-+},r_{+-},B_{-+}$ and $B_{+-},$ we get from
Eqs.(\ref{84},\ref{85}) and(\ref{88},\ref{89}) 
\begin{eqnarray}
x_{-+} &\equiv &\sin \gamma \sin \delta _{-+}=0.37\pm 0.21,\text{ }0.16\leq
x_{-+}\leq 0.58  \label{102} \\
x_{+-} &\equiv &\sin \gamma \sin \delta _{+-}=1.11\pm 0.68,\text{ }0.43\leq
x_{+-}\leq 1  \label{103} \\
z_{-+} &\equiv &\cos \gamma \cos \delta _{-+}=-0.03\pm 0.31,\text{ }%
-0.34\leq z_{-+}<0.28  \label{104} \\
z_{+-} &\equiv &\cos \gamma \cos \delta _{+-}=-0.03\pm 0.30,\text{ }%
-0.33\leq z_{+-}<0.27  \label{105}
\end{eqnarray}
From Eqs.(\ref{102}) and (\ref{103}), we obtain the following limits on the
final state strong phases, for $\gamma $ in the range $51^{\circ }\leq
\gamma \leq 75^{\circ }$\cite{8} 
\begin{eqnarray*}
10^{\circ } &\leq &\delta _{-+}\leq 48^{\circ }\text{ or }170^{\circ }\geq
\delta _{-+}\geq 132^{\circ } \\
27^{\circ } &\leq &\delta _{+-}\leq 90^{\circ }\text{ or }153^{\circ }\geq
\delta _{+-}\geq 90^{\circ }
\end{eqnarray*}
We note that $x$ and $z$ satisfy the following equations 
\begin{eqnarray}
\frac{x^{2}}{\sin ^{2}\gamma }+\frac{z^{2}}{\cos ^{2}\gamma } &=&1  \nonumber
\\
&&  \label{105a} \\
\frac{x^{2}}{\sin ^{2}\delta }+\frac{z^{2}}{\cos ^{2}\delta } &=&1  \nonumber
\end{eqnarray}
Figs(1,2) display these ellipses in ($x,z$) plane for $x_{-+},z_{-+}$ and $%
x_{+-},z_{+-}$ respectively for various values of $\gamma $ and $\delta .$
The experimental data restrict the allowed area in $(x,z)$ plane to the dark
shaded regions in Figs(1,2). It is clear from Fig 1, that for $%
z_{-+}<0,\gamma >65^{\circ },$ whereas $\gamma >70^{\circ }$ for $z_{-+}>0.$
From Figs 1 and 2, we also note that $\delta _{-+}\leq 40^{\circ }$ for $%
z_{-+}>0$ or $\delta _{-+}\geq 140^{\circ }$ for $z_{-+}<0$, whereas $\delta
_{+-}\geq 25^{\circ }$ or $\delta _{+-}\leq 155^{\circ }.$

Using the experimental values given in Eqs.(\ref{96}) and (\ref{97}), we
obtain 
\begin{eqnarray}
\frac{R_{-+}+R_{+-}}{R_{-+}^{\prime }+R_{+-}^{\prime }} &=&1.12\pm 0.22
\label{105b} \\
A_{CP}^{\prime } &=&0.14\pm 0.09  \label{105c} \\
C^{\prime }+A_{CP}^{\prime }\Delta C^{\prime } &=&-0.34\pm 0.17  \label{105d}
\end{eqnarray}
Since the experimental values for mixing induced asymmetries $S$ and $\Delta
S$ have large errors, there is no point in giving the numerical values for $%
S^{\prime }$ and $\Delta S^{\prime }.$

\textbf{To conclude:}

Using rotation in $SU(3)$ space, a set of relations between various decay
modes of $B$ mesons have been derived. In this way, we have avoided use of
group representation of $SU(3)$ symmetry. In particular we have related the
decay modes of $\bar{B}_{d}^{0}$ and $\bar{B}_{s}^{0}$ to $K\bar{K},K^{*}%
\bar{K},\bar{K}^{*}K$ and $\bar{B}_{s}^{0}\rightarrow K^{+}K^{-}$ to $\bar{B}%
_{d}^{0}\rightarrow \pi ^{+}K^{-}$. Due to lack of experimental data, some
of the relations cannot be tested. However for $B^{0}\rightarrow \rho
K,\omega K,\phi K$ decays we get 
\begin{eqnarray}
&&  \nonumber \\
\frac{\langle \Gamma \rangle _{\omega K}+\langle \Gamma \rangle _{\rho ^{0}K}%
}{\langle \Gamma \rangle _{\phi K}} &\approx &1,\text{ }\frac{S(\rho
^{0}K_{s})+S(\omega K_{s})}{2}=S(\phi K_{s})=-\sin 2\beta  \label{106} \\
C(\rho ^{0}K_{s}) &=&-C(\omega K_{s})  \label{107} \\
f_{\rho }-\frac{1}{3}f_{\omega }-\frac{2}{3}f_{\phi } &=&0  \label{108}
\end{eqnarray}
In more details, we have expressed the decay amplitudes $\left\langle
K^{*-}K^{+}\left| H_{w}(s)\right| \bar{B}_{s}^{0}\right\rangle =\left\langle
K^{*-}\pi ^{+}\left| H_{w}(s)\right| \bar{B}_{d}^{0}\right\rangle $ and $%
\left\langle K^{*+}K^{-}\left| H_{w}(s)\right| \bar{B}_{s}^{0}\right\rangle
=\left\langle \rho ^{+}K^{-}\left| H_{w}(s)\right| \bar{B}%
_{d}^{0}\right\rangle $ in terms of the decay parameters of $\bar{B}%
_{d}^{0}\rightarrow \rho ^{-}\pi ^{+}$ and $\bar{B}_{d}^{0}\rightarrow \rho
^{+}\pi ^{-}$ respectively. In particular in Eqs.(\ref{92}) and (\ref{93}),
we have obtained $r_{-+}(r_{+-})$ in terms of two unknown parameters $%
B_{-+}(B_{+-})$ as all other terms in these equations are known
experimentally. These equations follow from broken $SU(3).$ $SU(3)$ is a
good symmetry except for the masses and the matrix elements of weak currents
between vaccum and pseudoscalar or vector mesons belonging to octet or nonet
representations of $SU(3).$ Thus $SU(3)$ breaking effects are taken care of
by using physical values for masses and for $f_{\pi },f_{K},f_{\rho }$ and $%
f_{K^{*}}.$ Parameters $B_{-+}(B_{+-})$ are determined using factorization
for tree graph. Having determined $r_{-+}(r_{+-}),$ we obtain $%
z_{-+}(z_{+-}) $ and $x_{-+}(x_{+-})$ from Eqs.(\ref{88}), (\ref{89}), (\ref%
{84}) and (\ref{85}). From $x_{-+}$ and $x_{+-}$ following bounds on strong
phases are obtained 
\begin{equation}
10^{\circ }\leq \delta _{-+}\leq 48^{\circ },\text{ }27^{\circ }\leq \delta
_{+-}\leq 90^{\circ }  \label{109}
\end{equation}
For $z_{-+},z_{+-}<0,$ the angles lie in the second quadrant. Further from
Fig(1) , we get lower limit on $\gamma $ to be $65^{\circ },$ and for the
final state phases we get 
\begin{equation}
\delta _{-+}<40^{\circ }(\delta _{-+}\geq 160^{\circ })  \nonumber
\label{111}
\end{equation}
From Fig 2, we get 
\[
\delta _{+-}>25^{\circ }(\delta _{+-}<155^{\circ }) 
\]

\FRAME{ftbpFU}{286.0625pt}{219.9375pt}{0pt}{\Qcb{Plot of $x_{-+}$ versus $%
z_{-+}$ for various values of $\protect\gamma $ (solid curves) and $\protect%
\delta $ (dotted curves). The dark shaded area is the region allowed by
experimental data.}}{}{Figure }{\special{language "Scientific Word";type
"GRAPHIC";maintain-aspect-ratio TRUE;display "USEDEF";valid_file "T";width
286.0625pt;height 219.9375pt;depth 0pt;original-width
454.1875pt;original-height 346.125pt;cropleft "0";croptop "1";cropright
"1";cropbottom "0";tempfilename
'../../../../SWP25/Fayyazsahib/JCSCFK00.wmf';tempfile-properties "XP";}}

\FRAME{ftbpFU}{286.0625pt}{220.8125pt}{0pt}{\Qcb{Plot of $x_{+-}$ versus $%
z_{+-}$ for various values of $\protect\gamma $ (solid curves) and $\protect%
\delta $ (dotted curves\.{)} The dark shaded area is the region allowed by
experimental data.}}{}{Figure }{\special{language "Scientific Word";type
"GRAPHIC";maintain-aspect-ratio TRUE;display "USEDEF";valid_file "T";width
286.0625pt;height 220.8125pt;depth 0pt;original-width
456.875pt;original-height 346.125pt;cropleft "0";croptop "1";cropright
"1";cropbottom "0";tempfilename
'../../../../SWP25/Fayyazsahib/JCSCHE01.wmf';tempfile-properties "XP";}}

\textbf{Acknowledgments:}

The author acknowledges a research grant provided by the Higher Education
Commission of Pakistan to him as a Distinguished National Professor. The
author also wishes to thank Mr. Ijaz Ahmed for drawing the figures.

\textbf{Figure Captions:}

Figure1: Plot of $x_{-\,+}$ versus $z_{-\,+}$ for various values of $\gamma $
(solid curves) and $\delta $ (dotted curves). The dark shaded area is the
region allowed by experimental data.

Figure2: Plot of $x_{+\,-}$ versus $z_{+\,-}$ for various values of $\gamma $
(solid curves) and $\delta $ (dotted curves\.{)} The dark shaded area is the
region allowed by experimental data.


\begin{thebibliography}{9}
\bibitem{1} For recent reviews, see for instance,
M.Gronau,arXiv:hep-ph/0510153 v3, 2005; M.Gronau, E. Lunghi and D.Wyler,
arXiv:hep-ph/0410170 v3, 2004; C. W. Baur et al. arXiv: hep-ph/0401188 v2,
2004; M. Beneke and M. Neubert, Nucl. Phys. B 675 (2003) 33; U. Nierste,
,arXiv:hep-ph/0511125 v1, 2005; F.Forti, arXiv:hep-ex/0511010 v1, 2005; S.
Safir, Proc. Sci, HEP2005: 246 (2005), hep-ph/0512014, M.Gronau and
Jonathan.L.Rosner, hep-ph/0307095, M.Gronau and Jonathan.L.Rosner,
hep-ph/0610227.

\bibitem{2} See for example: Fayyazuddin and Riazuddin, A Modern
Introduction to Particle Physics Second Edition Page (134) World Scientific
(2000)

\bibitem{3} M.Gronau, Phys.Rev.D62 (2000) 014031.

\bibitem{4} N.G.Deshpande, Xiao-Gang He and Jian-Qing Shi, hep-ph/0002260
v2, Phys.Rev. D62 (2000) 034018

\bibitem{5} T.Das, V.S. Mathur and S.Okubo, Phys.Rev.Lett; 19 (1967) 470.
See also J.J. Sakurai, Currents and Mesons, The University of Chicago Press
(1969)

\bibitem{6} See for example: M.Beneke and M.Nuebert, Nucl.Phys B675,333(2003)

\bibitem{7} M.Gronau and J.Zupan, hep-ph/0407002, 2004;

\bibitem{8} Particle Data Group, W-M.Yao $\mathit{etal.}$ Journal of Physics
G \textbf{33},1 (2006)

\bibitem{9} Fayyazuddin hep-ph/0604008 v3, 2007.
\end{thebibliography}
\end{document}